\documentclass[prd,nofootinbib]{revtex4-1}
\usepackage[T1]{fontenc}
\usepackage[latin9]{inputenc}
\usepackage{amsmath}
\usepackage{amssymb}
\usepackage{graphicx,epsfig}
\begin{document}

\begin{flushright} 
KCL-PH-TH/2013-13, LCTS/2013-07, CERN-PH-TH/2013-079
\end{flushright} 

\vspace{1.5cm}

\title{Environmental CPT Violation in an Expanding Universe in String Theory
~\\
~\\
~\\} 

\author{John Ellis and Nick E. Mavromatos}

\affiliation{Theoretical Particle Physics and Cosmology Group, Department of Physics, King's College London,  Strand, London WC2R 2LS, UK. and \\
CERN, Theory Division, Physics Department, Geneva 23 CH-1211, Switzerland. }

\author{Sarben Sarkar}

\affiliation{Theoretical Particle Physics and Cosmology Group, Department of Physics, King's College London,  Strand, London WC2R 2LS, UK.}

\begin{abstract}
~\\
~\\
~\\
We consider a model of an expanding Universe in string theory that yields  
`environmental'  CPT violation for fermions, in the sense of different dispersion 
relations for fermions and antifermions. These are induced 
by a cosmological background with constant torsion provided by the 
Kalb-Ramond antisymmetric tensor field (axion) of the string gravitational multiplet. This effect induces 
different densities of neutrinos and antineutrinos  while in chemical equilibrium, 
offering new scenarios for leptogenesis and baryogenesis even in the absence of CP violation.
~\\
~\\
~\\
~\\
~\\
April 2013

\end{abstract}

 \maketitle

\section{Introduction and Summary}

The visible matter in the Universe is composed overwhelmingly of matter particles, with no 
detectable concentrations of anti-matter
particles. However, according to the standard Big Bang theory in the absence of a primordial matter-antimatter
asymmetry, matter and antimatter would have been present in equal amounts in the early
radiation-dominated Universe. The observation of charge-parity (CP) violation in particle
physics~\cite{Christenson:1964fg} prompted Andrei Sakharov~\cite{Sakharov:1967dj}
to conjecture that fundamental particle interactions could have generated a Baryon Asymmetry in the Universe (BAU), if
the following conditions could all be met:
\begin{itemize}
\item
Baryon-number-violating interactions that allow the generation of states with $B\neq 0$ starting from an initial
state with $B =0$, where $B$ is the baryon number;
\item
Interactions capable of distinguishing between matter and antimatter - assuming CPT symmetry, this would
require violation of both C and CP;
\item
Since matter-antimatter asymmetry is impossible in chemical equilibrium, one also requires
some breakdown of chemical equilibrium during an epoch in the early Universe.
\end{itemize}

The asymmetry between the baryon ($n_B$) and anti-baryon ($n_{\overline{B}}$) number densities 
observed today is estimated to require~\cite{Gamow:1946eb}
\begin{equation}
\Delta n = \frac{n_{B}-n_{\overline{B}}}{n_{B}+n_{\overline{B}}}\sim\frac{n_{B}-n_{\overline{B}}}{s}=(8.4-8.9)\times10^{-11}\label{basym}
\end{equation}
at the early stages of the expansion for times $t \sim 10^{-6}$~s
and temperatures $T \sim 1$~GeV, where $s$ denotes the
entropy density. The CP violation observed~\cite{Christenson:1964fg} and predicted within the
Standard Model (SM) of particle physics could not have induced
an asymmetry as large as (\ref{basym})~\cite{Kuzmin:1985mm}.
There are several ideas that go beyond the SM, e.g.,  grand unified
theories (GUTs), supersymmetry, extra dimensions and 
the decays of massive sterile neutrinos, that provide additional sources of CP violation and hence
scenarios for generating the BAU: see~\cite{Shaposhnikov:2009zz,Shaposhnikov:2006xi,Lindner:2010wr,Kusenko:2010ik,
Randall:1999ee,Merle:2011yv,Barry:2011wb} for recent examples.

A basic assumption in the Sakharov scenario is that CPT
symmetry~\cite{cpttheorem}  (where $T$ denotes the time-reversal operation) holds in the very early Universe.
Indeed, CPT invariance
is intrinsic to all known \emph{local} effective \emph{relativistic}
field theories without gravity, such as those upon which current particle-physics phenomenology
is based. However, CPT may be violated, e.g., as the result of a breakdown of Lorentz symmetry,
as in the Standard Model Extension of~\cite{kostelecky} and in models of quantum 
gravity backgrounds~\cite{Barenboim:2004ev,Barenboim:2004wu,Mavromatos:2010pk}, and the requirement
of a breakdown of equilibrium can be dropped if the
requirement of CPT is relaxed~\cite{Bertolami:1996cq}. 
One attractive scenario is that the violation of CPT (denoted by CPTV) early in the expansion
of the Universe, might have generated a primordial
lepton asymmetry, after which either
sphaleron processes~\cite{PhysRevD.36.581} or Baryon-Lepton (B-L) conserving processes
in some GUT communicated this lepton asymmetry
to the baryon sector; in such a scenario, this mechanism which we call  CPTV leptogenesis, produces the observed BAU.
 CPTV in the early Universe would thus
have obviated the need for extra sources of CP violation, such as sterile neutrinos 
and/or supersymmetry, in order to obtain the observed BAU. 

In this paper we  present a new model for CPTV leptogenesis, 
proposed in a preliminary form in~\cite{mscptv}. Our approach follows broadly an earlier framework 
proposed in~\cite{Lambiase:2006md, Debnath:2005wk}, but differs crucially in that the full gravitational 
multiplet \cite{Zwiebach:789942} arising in string theory is used. The model is formulated in the Einstein
frame, and is based on a variant of a 
prototype for an expanding Universe proposed in~\cite{aben}, which is an exact solution to the equations
of motion of the effective field theory derived from the stringy world-sheet $\sigma$ model. We extend it 
here to consider the effects on fermion propagation in a background with constant non-zero torsion~\footnote{See~\cite{AMT1,AMT2} for
previous examples of non-trivial cosmological sigma models that include antisymmetric tensor fields and produce 
backgrounds with constant axial $B^0$ vectors, which are analogous to the linear dilaton background discussed here.}.
provided by the antisymmetric Kalb-Ramond tensor in the gravitational multiplet of the string. 
The leptogenesis in this model is due to torsion-induced CPTV in the dispersion relations of fermions
and antifermions. Although the model presented here is a toy prototype, 
it contains some nontrivial features that may survive the extension to more realistic and detailed string cosmologies. 

The structure of this paper is as follows: in 
Section~\ref{sec:string} we discuss the string model of~\cite{aben} and the underlying world-sheet conformal field theory 
formalism that leads to a non-trivial torsion background, and we explain how such a background leads to
CPTV dispersion relations for fermions and antifermions. We then discuss how this would lead to 
different number densities for neutrinos and antineutrinos, providing a possible scenario for 
leptogenesis without the need for any extra sources of CP violation. Finally, an
outlook is presented in Section~\ref{sec:conc}.

\section{Environmental CPT violation in a String Cosmological Background Geometry with Torsion}\label{sec:string}

The massless gravitational multiplet in string theory contains the graviton, described by a
spin-2 symmetric tensor $g_{\mu\nu}$, a spin-0 scalar dilaton, $\Phi$,
and an antisymmetric tensor, $B_{\mu\nu}$. This Kalb-Ramond field appears in the string effective action only through its 
totally-antisymmetric field strength, $H_{\mu\nu\rho} = \partial_{\left[ \mu \right.} B_{\left.\nu\rho\right]}$, 
where  $[ \dots ]$ denotes antisymmetrization of the indices within the brackets. The calculation of  
string amplitudes~\cite{sloan} shows that $H_{\mu\nu\rho}$ plays the role of \emph{torsion} in a generalized connection.
In the Einstein frame, to $O(\alpha^\prime)$ the four-dimensional bosonic part of the  effective action of the string is:
\begin{equation}\label{ea}
S= \frac{M_s^2 \, V^{c}}{16\pi} \int d^4x \sqrt{-g} \Big( R(g) - 2 \partial^\mu \Phi \partial_\mu \Phi - \frac{1}{12} e^{-4\Phi} 
H_{\mu\nu\rho} \, H^{\mu\nu\rho} + \dots \Big) \, ,
\end{equation}
where $M_s=1/\sqrt{\alpha^\prime}$ is the string mass scale, the compactification volume $V^{(c)}$
and the compact radii are expressed dimensionlessly in units of $\sqrt{\alpha'}$. It can be shown that
the term in $S$  that is the square of the field strength can be combined with the Einstein scalar curvature term $R(g)$ to form
a generalized curvature term $\overline{R}(g, \overline{\Gamma })$. This generaliszed curvature is defined in terms of
a generalized Christoffel symbol (connection) 
$\overline{\Gamma}$:
\begin{equation}\label{generalised}
\overline{\Gamma}^\lambda_{\,\,\,\mu\nu} = \Gamma^\lambda_{\,\,\,\mu\nu} + e^{-2\Phi} H^\lambda_{\mu\nu} \equiv
\Gamma^\lambda_{\,\,\,\mu\nu} + T^\lambda_{\,\,\,\mu\nu} \, ,
\end{equation}
where $\Gamma^\lambda_{\,\,\,\,\mu\nu} = \Gamma^\lambda_{\,\,\,\,\nu\mu} $ is the torsion-free Einstein-metric connection, 
and $T^\lambda_{\,\,\, \mu\nu} = - T^\lambda_{\,\,\,\nu\mu}$ is the torsion. 

Using general covariance,
the four-dimensional Lagrangian ${\mathcal L}_f $ for a spin-1/2 Dirac fermion $\psi$ in this background with torsion is 
(up to constants of proportionality):
\begin{equation}\label{fermion}
{\mathcal L}_f  \sim \sqrt{-g}\, \Big( i \overline{\psi} \, \gamma^a D_a \, \psi - m \overline{\psi} \, \psi \Big) \, : 
\quad D_a = \partial_a - \frac{i}{4} \overline{\omega}_{bca}\, \sigma^{bc} ~, \quad \sigma^{ab} = \frac{i}{2} [ \gamma^a\, , \, \gamma^b ] \, ,
\end{equation}
where Latin indices refer to tangent space, and $\omega$, the generalized spin connection (with torsion), is:
\begin{equation}\label{Htors}
\overline{\omega}_{bcd} = e_{b\lambda} \Big( \partial_a \, e^\lambda_{\, a}  + \overline{\Gamma}_{\gamma\, \mu}^\lambda \, 
e^\gamma_{\, c} \, e^\mu_{\, a} \Big) ~.
\end{equation}
Using properties of the Dirac $\gamma$-matrices, the Lagrangian (\ref{fermion}) can be cast in the form 
\begin{equation}\label{Bvector}
\mathcal{L}=\sqrt{-g}\,\overline{\psi}\left(i\gamma^{a}\partial_{a}-m+\gamma^{a}\gamma^{5}B_{a}\right)\psi~,\quad B^{d}
=\epsilon^{abcd}e_{b\lambda}\left(\partial_{a}e_{\,\, c}^{\lambda}+\Gamma_{\nu\mu}^{\lambda}\, e_{\,\, c}^{\nu}\, e_{\,\, a}^{\mu}\right)~.
\end{equation}
The space-time curvature background therefore has the effect of
inducing an `axial' background field $B_{a}$ that is known to be non-trivial
in certain anisotropic space-time geometries, such as Bianchi-type
cosmologies or in regions of space-time near rotating (Kerr) primordial black 
holes~\cite{Debnath:2005wk,Mukhopadhyay:2005gb,Mukhopadhyay:2007vca,Sinha:2007uh}. 

For the application to particle-antiparticle asymmetry, we are
interested in an axial field $B_{a}$ that is constant in some local frame,
in which case CPT is violated  `environmentally' as a consequence of background-induced Lorentz violation,
since the dispersion relations of fermions and antifermions differ in such
backgrounds~\footnote{We note at this stage that in the case of a constant background $B^a$ 
the Lagrangian (\ref{Bvector}) falls into the framework of the 
Standard Model Extension~\cite{kostelecky}, with the Lorentz- and CPT-violating `axial' parameter $b_\mu$ of that 
formalism being identified with the background field $B^a$. String backgrounds of different type that violate 
Lorentz Invariance and CPT symmetry have also been discussed in this context 
explicitly in~\cite{koststring1,koststring2,koststring3}.}. Explicitly we have~\cite{Debnath:2005wk,Mukhopadhyay:2005gb,Mukhopadhyay:2007vca,Sinha:2007uh}.
\begin{equation}
E=\sqrt{(\vec{p}-\vec{B})^{2}+m^{2}}+B_{0}~,\quad{\overline{E}}=\sqrt{(\vec{p}+\vec{B})^{2}+m^{2}}-B_{0}~.\label{nunubardr}
\end{equation}
Similar considerations apply to (left-handed) Majorana fermions, in which case there is a
difference between the dispersion relations for left-handed spinors and their conjugate (right-handed) spinors.
In this case, the coupling of the (left-handed) Majorana 
fermions to the background four-vector $B^a$ has the following
form in the Weyl representation~\cite{Mukhopadhyay:2005gb,Mukhopadhyay:2007vca,Sinha:2007uh}.
\begin{equation}
{\mathcal L}_{\rm major} \ni \sqrt{-g} \, \Big( \overline{\psi}^c_L \gamma^a \psi^c_L - \overline{\psi}_L \gamma^a \psi_L \Big) \, B_a~,
\end{equation}
which implies different (CPTV) dispersion relations of the form (\ref{nunubardr}) for
(left-handed) Majorana spinors and their (right-handed) conjugates.

The existence of a frame in which the four-vector $B^a$ is constant in space-time
has not been demonstrated in the literature existing so far. 
However, as argued in~\cite{mscptv} and shown below,
this is the case in the string background we consider here, in which the torsion provides a space-time-independent 
$B^0$ and vanishing $\vec B$. 
The torsion $T_{\mu\nu\rho}$ associated with the generalized connection $\overline{\Gamma}_{\mu\nu}^\rho$
(\ref{generalised})
yields \emph{non-zero} contributions to the fermion interactions in (\ref{Bvector}),
with $ e_{b\lambda}\, \overline{\Gamma}_{\gamma\, \mu}^\lambda \, e^\gamma_{\, c} \, e^\mu_{\, a}$. 
Even in the case of constant vierbeins, i.e., in the
limit of flat Minkowski space-time, there are non-trivial interactions between the fermions and the torsion part of the spin-connection.

Solutions to the conformal invariance conditions of the low-energy 
string effective action that are exact to all orders in $\alpha^\prime$ have been presented in~\cite{aben}. 
The antisymmetric tensor field strength in the four `large' (uncompactified) dimensions of the string
can be written uniquely as
\begin{equation}\label{Hfield}
H_{\mu\nu\rho} = e^{2\Phi} \epsilon_{\mu\nu\rho\sigma} \partial^\sigma b (x) \, ,
\end{equation}
where $\epsilon_{0123} = \sqrt{g}$ and $\epsilon^{\mu\nu\rho\sigma} = |g|^{-1} \epsilon_{\mu\nu\rho\sigma}$, 
with $g$ the metric determinant.  The field
$b(x)$ is a pseudoscalar axion-like field. The dilaton $\Phi$ and axion $b$ fields
appear as Goldstone bosons of spontaneously broken scale symmetries of the string vacuum, 
and so are exactly massless classically. Such fields appear in the effective action only through their derivatives.
In the exact solution of~\cite{aben} in the string frame both the dilaton and axion fields are linear in the 
target time $X^0$ in the $\sigma$-model (Jordan) frame: $\Phi (X^0), b(X^0) \sim X^0$. 
This solution shifts the minima of all the fields in the effective action that couple to the dilaton and axion by an 
amount that is independent of space and time.

In the Einstein frame that is relevant for cosmological observations, the temporal component of the
metric is normalized to $g_{00} =+1$ by an appropriate change of the time coordinate. In this setting,
the solution of~\cite{aben} leads to a Friedmann-Robertson-Walker (FRW) metric with scale factor $a(t) \sim t$, 
where $t$ is the FRW cosmic time and the dilaton field  $\Phi$ behaves as 
\begin{equation}\label{dilaton}
\Phi(t) = - {\rm ln} t + \phi_0 ~, 
\end{equation}
where $\phi_0$ is a constant, and the axion field $b(x)$ is  linear in $t$, cf, Eq. (\ref{axion}) below. 
There is an underlying world-sheet conformal field theory with central charge 
\begin{equation}\label{ccharge}
c = 4 - 12 Q^2 - \frac{6}{n + 2} + c_I~,
\end{equation}
where $Q^2 (> 0 )$ is the central-charge deficit and  $c_I$ is the central charge associated with the 
world-sheet conformal field theory of the compact `internal' dimensions of the string model~\cite{aben}. 
The world-sheet ghosts must cancel because of the requirement of reparametrization invariance of the
world-sheet co-ordinates entails that $c=26$. 

The solution for the axion field is
\begin{equation}\label{axion}
b(x) = \sqrt{2} e^{-\phi_0} \, \sqrt{Q^2} \,  \frac{M_s}{\sqrt{n}} t~,
\end{equation}
where $M_s$ is the string mass scale and $n$ is a positive integer, associated with the level of the Kac-Moody 
algebra of the underlying world-sheet conformal field theory. For non-zero $Q^2 $ there is an additional  dark energy 
term in  the effective target space-time action of the string~\cite{aben} of the form
$\int d^4 x \sqrt{-g} e^{2\Phi} (- 4Q^2)/\alpha^\prime $.
The linear axion field (\ref{axion}) remains a non-trivial solution even in the limit of a static space-time
with a constant dilaton field~\cite{aben}.  In this case the space-time is an Einstein universe with a
positive cosmological constant and constant positive curvature proportional to
$6/(n+2)$. For the solutions of~\cite{aben}, the covariant torsion tensor  $e^{-2\Phi} H_{\mu\nu\rho} $ is constant,
as can be seen from (\ref{generalised}) and (\ref{Hfield},
since the exponential  dilaton factors cancel out in the relevant expressions. Only the spatial components of the torsion 
are nonzero in this case:
$T_{ijk} \sim \epsilon_{ijk} {\dot b} = \epsilon_{ijk} \sqrt{2 Q^2} e^{-\phi_0} \,  \frac{M_s}{\sqrt{n}},$
where the dot denotes a derivative with respect to $t$.  

On the basis of (\ref{generalised}), (\ref{Hfield}) and (\ref{Bvector}), we  observe that
only the temporal component $B^0$  of the vector $B^d$ is non-zero 
and constant, of order~\footnote{We note that the torsion-free gravitational part of the connection (for the FRW or flat case)
yields a vanishing contribution to $B^0$.}
\begin{equation}\label{b0string}
B^0 = \epsilon^{ijk} T_{ijk} = 6\,  \sqrt{2 Q^2} e^{-\phi_0} \, \frac{M_s}{\sqrt{n}} ~ {\rm GeV} > 0 \, .
\end{equation}
One obtains from the dispersion relations (\ref{nunubardr}) different populations of fermions and antifermions while
in thermal equilibrium, as a result of the following expressions for the
particle and antiparticle  distribution functions~\footnote{We note that the torsion and the other background fields (metric and dilaton) 
in the $\sigma$-model effective Lagrangian have only gravitational-strength interactions. The rates for these interactions are slower 
than the expansion of the Universe, so they are not in thermal equilibrium and the underlying conformal field theory is
unaffected. It is only the matter sector that has finite-temperature corrections.}:
\begin{equation}
f(E,\mu)=[{\rm exp}(E(\vec p)-\mu)/T)\pm1]^{-1}~, \quad 
f(\overline{E(\vec p)},\bar{\mu})=[{\rm exp}(\bar{E}(\vec p)-\bar{\mu})/T)\pm1]^{-1}~,\label{cptvf}
\end{equation}
where $\vec{p}$ is the $3-$momentum and $\mu$ the chemical potential. Our convention is that
an overline over a quantity refers to an antiparticle, the $+$ sign applies for a fermionic (anti-)particle 
and the $-$ applies for a bosonic (anti-)particle~\footnote{
We note that in the discussion of \cite{aben}, it was observed that  the central charge deficit $Q^2 > 0$ implies tachyonic (i.e. negative) shifts in the 
mass squared  of the boson fields, while fermion masses were not affected. We see in this work that the presence of antisymmetric torsion does affect the dispersion relations (\ref{nunubardr}) for fermions .}. 

Since the torsion couples universally to all fermonic species, the above considerations apply to all flavours of 
fermions, including both quarks and leptons. In the following, we first concentrate on the case of 
light (left-handed) neutrinos, with a view to their potential importance in leptogenesis. In this case, we postulate an early
epoch during which lepton-number-violating processes are sufficiently rapid to maintain
chemical equilibrium for neutrinos and antineutrinos.
As we see from Eqs.~(\ref{b0string}), (\ref{nunubardr}) and (\ref{cptvf}),
one obtains when $B^0 \ll T$~\cite{mscptv,Debnath:2005wk,Mukhopadhyay:2005gb,Mukhopadhyay:2007vca,Sinha:2007uh}:
\begin{equation}
\Delta n_{\nu}\equiv n_{\nu}-n_{\overline{\nu}}\sim g^{\star}\, T^{3}\left(\frac{B_{0}}{T}\right)\label{bianchi}
\end{equation}
where $g^{\star}$ is the number of degrees of freedom for the (relativistic) 
neutrino, and an excess of particles over antiparticles is predicted
when $B_{0}>0$. If $B^0$ is independent of temperature, as assumed but not demonstrated in the models 
of~\cite{Debnath:2005wk,Mukhopadhyay:2005gb,Mukhopadhyay:2007vca,Sinha:2007uh}, the 
CPTV asymmetry induced by the background decreases with the temperature. 

At temperatures $T<T_{d}$, where $T_{d}$ denotes the temperature
at which lepton-number violating processes decouple, which depends on the details of the underlying model, 
the ratio of the net lepton
number $\Delta L$ (neutrino asymmetry) to entropy density (which
scales as $T^{3}$) remains constant,
\begin{equation}
\Delta L(T<T_{d})=\frac{\Delta n_{\nu}}{s}\sim\frac{B_{0}}{T_{d}} \; ,
\label{dlbianchi}
\end{equation}
As in other leptogenesis scenarios, this lepton asymmetry can then be communicated
to the baryon sector to produce the observed baryon asymmetry via the
B-L-conserving sphaleron processes of the standard model or B-L-conserving processes  in the context of 
some grand unified theory. It should be emphasized, however, that unlike conventional
leptogenesis scenarios there is no need to postulate any CP violation in any lepton-number-violating
processes. The sign of $B_0$ determines whether there is an abundance of neutrinos over antineutrinos or 
vice versa. For successful baryogenesis  in a B-L conserving scenario, e.g.,
exploiting sphalerons in the Standard Model, we need $B_0 > 0$. 

We consider first the case of massive Majorana neutrinos, in which case torsion induces oscillations 
between neutrinos and antineutrinos, as in the original suggestion of Pontecorvo~\cite{Pontecorvo2,Bilenky:1978nj}.
These oscillations are induced by the mixing of neutrino and antineutrino states to produce mass eigenstates
due to the constant `environmental' field
$B^0$~\cite{Mukhopadhyay:2007vca,Sinha:2007uh}. To see this, we
consider the Lagrangian for Majorana neutrinos in the presence of $B_a$,
written in terms of two-component (Weyl) spinors $\psi, \psi^c$ 
(a generic four-component Majorana spinor $\Psi$ may be written in our notation as 
$\Psi = \begin{pmatrix} \psi^c_L \\ \psi_L \end{pmatrix}$, 
where from now on we omit the left-handed suffix $L$):
\begin{equation}\label{nulagr}
{\mathcal L}_{\nu} = \sqrt{-g} \Big[ \big({\psi^c}^\dagger \quad \psi^\dagger \big) \frac{i}{2} \gamma^0 \, \gamma^\mu \, D_\mu \begin{pmatrix} \psi^c \\ \psi \end{pmatrix} - \big({\psi^c}^\dagger \quad \psi^\dagger \big) \begin{pmatrix} -B_0 \quad -m \\ -m \quad B_0 \end{pmatrix} \, \begin{pmatrix} \psi^c \\ \psi \end{pmatrix} \, ,
\end{equation}
where we assume for now that the neutrino has only lepton-number-violating Majorana-type masses~\footnote{The 
extension to Dirac-type masses involving sterile neutrino will be discussed later on.}.
We see from (\ref{nulagr}) that, in the presence of torsion,
there are non-trivial and unequal diagonal lepton-number-conserving
entries in the mass matrix $\mathcal{M}$ for $\psi$ and $\psi^c$: 
\begin{equation}\label{mixing} 
{\mathcal M} = \begin{pmatrix} - B_0 \quad - m \\ - m \quad B_0 \end{pmatrix}~.
\end{equation}
The mass matrix (\ref{mixing}) is hermitean, so can be diagonalised by a unitary matrix, 
leading to two-component mass eigenstates $| \chi_{i,j} \rangle$ that are mixtures of the 
energy eigenstates $|\psi \rangle$ and $|\psi^c \rangle$:
\begin{eqnarray}\label{masseig}
| \chi_1 \rangle  & = & {\mathcal N}^{-1} \, \{ \Big( B_0 + \sqrt{B_0^2 + m^2} \Big)\, | \psi^c \rangle + m \, |\psi \rangle \} ~,  \nonumber \\
| \chi_2 \rangle  & = & {\mathcal N}^{-1} \, \{ - m \, |\psi^c \rangle + \Big( B_0 + \sqrt{B_0^2 + m^2} \Big)\, | \psi  \rangle  \} ~,  
\end{eqnarray}
where
\begin{equation}
{\mathcal N} \equiv \Big[ 2 \Big(B_0^2 + m^2 + B_0 \sqrt{B_0^2 + m^2} \Big)\Big]^{1/2} ~,
\label{norm}
\end{equation}
with mass eigenvalues 
\begin{equation}\label{masses}
m_{1,2} = \mp \sqrt{B_0^2 + m^2} ~.
\end{equation}
Notice that in the presence of $B_0$ the mass eigenstates are \emph{different} from the energy eigenstates of the Hamiltonian.

The above mixing can be expressed by writing the four-component neutrino spinor in terms of
$\psi$ and $\psi^c$ using an angle $\theta$~\cite{Sinha:2007uh}:
\begin{equation}\label{mixingangle}
\nu \; \equiv \; \begin{pmatrix} \chi_1 \\ \chi_2 \end{pmatrix} = \begin{pmatrix} {\rm cos}\, \theta  \quad {\rm sin}\, \theta \\ -{\rm sin}\, \theta
\quad  {\rm cos} \, \theta \end{pmatrix} \, \begin{pmatrix} \psi^c \\ \psi \end{pmatrix}~: \qquad {\rm tan} \, \theta \equiv \frac{m}{B_0 + \sqrt{B_0^2 + m^2}}~.
\end{equation}
It is readily seen that the four-component spinor $\nu$ is also Majorana, 
as it satisfies the Majorana condition $\nu^c = \nu$. We note that in the absence of torsion, $B_0 \to 0$,
the mixing angle between the two-component spinors $\psi$ and $\psi^c$ is maximal: $\theta = \pi/4$,
whereas it is non-maximal when $B_0 \ne 0$.

The mixing (\ref{mixingangle}) enables us to understand the difference between the densities of
fermions and antifermions mentioned earlier (\ref{bianchi}).
The expectation values of the number operators of $\chi_{i} , i=1,2$ in energy eigenstates are given by:
\begin{eqnarray}
N_{\chi_1} & = & < : \chi_1^\dagger \, \chi_1: > =  {\rm cos}^2 \theta \, < : {\psi^c}^\dagger \, \psi^c : >  + {\rm sin}^2 \theta \, < : {\psi }^\dagger \, \psi : >  ~, \nonumber  \\
N_{\chi_2} & = & < :\chi_2^\dagger \, \chi_2: > =  {\rm sin}^2 \theta \, < : {\psi^c}^\dagger \, \psi^c : >  + {\rm cos}^2 \theta \, < : {\psi }^\dagger \, \psi : >  ~, 
\end{eqnarray}
where cross-terms do not contribute. We observe that, for general $\theta \ne \pi/4 $, \emph{i.e}., $B_0 \ne 0$,
as seen in (\ref{mixingangle}), there is a difference between the populations of $\chi_{1}$ and $\chi_{2}$:
\begin{equation}\label{n12}
N_{\chi_1} - N_{\chi_2} = {\rm cos}\, 2\theta \Big( <n_{\psi^c} > -  <n_{\psi} > \Big)~,
\end{equation}
where $<n_{\psi}> = <: \psi^\dagger \, \psi :> \ne <n_{\psi^c}> = <: {\psi^c}^\dagger \, \psi^c :>$ are the corresponding number operators for the energy eigenstates.
 
This difference in the neutrino and antineutrino populations (\ref{bianchi})
is made possible by the presence of fermion-number-violating fermion-antifermion oscillations,
whose probability was calculated in~\cite{Sinha:2007uh}: 
\begin{equation}\label{probmix}
{\mathcal P}(t) = |\langle \nu_1 (t) | \nu_2(0) \rangle |^2 = {\rm sin}^2 \theta  \, {\rm sin}^2 \Big(\frac{E_\nu - E_{\nu^c}}{2} \, t \, \Big) = \frac{m^2}{B_0^2 + m^2} \, {\rm sin}^2 (B_0 \, t)~,
\end{equation}
where we used  (\ref{nunubardr}) with $\vec B = 0$, as in our specific background,
and the definition of the mixing angle (\ref{mixingangle}). 
The time evolution of the system is calculated using $|\psi \rangle$ and $|\psi^c\rangle$, which are eigenstates of the Hamiltonian.

In the case of relativistic neutrinos moving close to the speed of light, the oscillation length obtained from (\ref{probmix}) is 
\begin{equation}\label{osclength}
L = \frac{\pi\, \hbar \, c }{|B_0|} = \frac{6.3 \times 10^{-14}\, {\rm GeV} }{B_0}~{\rm cm}.
\end{equation}
where we have reinstated $\hbar$ and $c$, and $B_0$ is measured in GeV.
For oscillations to be effective at any given epoch in the early Universe, 
this length has to be less than the size of the Hubble horizon.
We assume that a cosmological solution of the form discussed in~\cite{aben}, with a scale factor increasing linearly with time,
is applicable some time after any earlier inflationary epoch. For a temperature
$T_d \sim 10^{9}$~GeV, the relevant Hubble horizon size $\sim 10^{-12}$~cm.
On the other hand, we see from (\ref{dlbianchi}) that the correct order of magnitude for the lepton
asymmetry $\sim 10^{-10}$ is obtained if $B_0 \sim 10^{-1}$ GeV.
For this value of $B_0$, the oscillation length (\ref{osclength}) $\sim 6.3 \times 10^{-13}$~cm,
which is  within the Hubble horizon size $\sim 10^{-12}$~cm. 
This implies that neutrino/antineutrino oscillations occur sufficiently rapidly to establish
chemical equilibrium and hence a lepton asymmetry~\footnote{Since the coupling of fermions to torsion is universal, 
the background would also couple to quarks and charged leptons. 
However, the splitting of a four-component spinor into a pair of two-component ones, and the subsequent
derivation of the above results on oscillations between fermions and antifermions, 
is valid exclusively for neutrinos and cannot apply to quarks or charged leptons.
Fermion-antifermion oscillations for these species would be forbidden by charge conservation.}.

In the above we considered a single generation of Majorana neutrino, 
without a Dirac mass term, $m_D$, which is required in seesaw mechanisms for the generation
of light neutrino masses. The inclusion of a Dirac mass term modifies the mass matrix (\ref{mixing})
via additional contributions to the diagonal mass terms that
have the same sign for the two-component neutrino $\psi$ and its conjugate spinor $\psi^c$:
\begin{equation}\label{mixingdirac} 
{\mathcal M} = \begin{pmatrix} m_D - B_0 \qquad \,\, - m \\ - m \qquad \,\, m_D + B_0 \end{pmatrix}~,
\end{equation}
which modifies the mass eigenvalues (\ref{masses}) to 
\begin{equation}\label{massesdirac}
m^{\rm s}_{1,2} = m_D \mp \sqrt{B_0^2 + m^2} ~.
\end{equation}
In this case, following the above steps and setting $|\vec p| \simeq E$, the average energy of an ultra-relativistic neutrino, 
one can recalculate the oscillation probability and oscillation length. For ultra-relativistic neutrinos $p \gg m$,
the oscillation probability becomes~\cite{Sinha:2007uh}
\begin{equation}
{\mathcal P}^s(t) \simeq \frac{m^2}{B_0^2 + m^2} {\rm sin}^2 \Big(\frac{m_D \, \sqrt{B_0^2 + m^2}}{E} \, t \Big)~, 
\label{Diracoscillations}
\end{equation}
leading to an energy-dependent oscillation length:
\begin{equation}
L^s \simeq \frac{\pi E}{m_D \, \sqrt{B_0^2 + m^2}}~. 
\end{equation} 
Flavour mixing in the presence of a background $B_0$ can also be considered, with effects on flavour oscillations
that fall beyond the scope of our discussion here. We refer the reader to ~\cite{Sinha:2007uh}, where flavour oscillations in the presence of background $B_0$ have been studied in some detail. 

\section{Outlook \label{sec:conc} }

As we have seen, the stringy cosmology proposed in~\cite{aben} introduces the possibility of a constant
torsion background in which a neutrino-antineutrino asymmetry develops as long as $B_0 \ne 0$.
Clearly, more work is needed to flesh out the details of a possible leptogenesis scenario exploiting this CPTV
mechanism, but we conclude by discussing some of the open issues.

It is clear that $B_0$ is negligible in the present-day Universe, which is characterized by a Minkowski space-time background.
A phase transition directly from the value of $B_0 \neq 0$ during leptogenesis to $B_0 = 0$ is the simplest possibility,
on the other hand in~\cite{aben} a series of phase transitions was envisaged, 
in which the central charge deficit of the model $Q^2$ gradually reduced step-wise to zero.
At each phase transition the value of the  $H$-torsion background would change from a larger to a smaller value.
In this approach, conformal field theories with different central charges characterize different epochs of the early 
Universe~\cite{aben}.

In the simplest model, in which
there is a single phase transition at a temperature $T_d$ at which $B_0$ switches off, 
this temperature would correspond to a decoupling temperature for the fermion-number-violating 
torsion-induced interactions in this model. The lepton number violation would freeze out,
leaving a lepton asymmetry (\ref{dlbianchi}) that would be converted partially into a baryon asymmetry
via sphaleron transitions in the Standard Model, as in other leptogenesis scenarios. 
In terms of the underlying neutrino-antineutrino oscillations, $T_d$ would also
be also the temperature at which such oscillations stop. We emphasize, however, that, unlike
other leptogenesis scenarios, no CP violation is necessary.
More complicated scenarios could also be envisaged, in which the non-zero torsion relaxes
to zero in a series of discrete steps, as originally proposed in~\cite{aben}, but this is not
essential for our purposes.

States with different values of $B_0$ correspond in the scenario of~\cite{aben} 
to different conformal (fixed) points in moduli space. Transitions between such states correspond to 
phase transitions in the early universe that can be described by
non-conformal (Liouville) time evolution, as discussed in~\cite{nickliouv}.
A fuller study of the scenario for leptogenesis described here should
include a discussion of such transitions, which lies beyond the scope of this
exploratory work.

\section*{Acknowledgments}

The work of J.E. and N.E.M. was supported in part by the London Centre for
Terauniverse Studies (LCTS), using funding from the European Research
Council via the Advanced Investigator Grant 267352, and 
by the UK STFC under the research grant ST/J002798/1.

\section*{References}

\addcontentsline{toc}{chapter}{{\bf Bibliography}}
\bibliography{MSCPTVUniv2}
\bibliographystyle{apsrev4-1}

\end{document}